\documentclass[prd,aps,showpacs,tightenlines, 10pt, reprint,nofootinbib,superscriptaddress,floatfix,twocolumn]{revtex4-2}

\usepackage[utf8]{inputenc} 
\usepackage{acronym}
\usepackage{amsmath}
\usepackage{amsthm}  
\usepackage{amssymb}
\usepackage{mathtools}
\usepackage{phaistos}
\usepackage{graphicx}
\usepackage[section]{placeins}
\usepackage{orcidlink}
\usepackage{enumerate}

\usepackage{physics}
\usepackage{IEEEtrantools}  
\usepackage{xspace}
\usepackage{cancel}
\usepackage[normalem]{ulem} 

\usepackage{array}
\newcolumntype{C}[1]{>{\centering\let\newline\\\arraybackslash\hspace{0pt}}m{#1}}

\usepackage{cellspace}
\usepackage{comment}
\usepackage{epsfig}
\usepackage{multirow} 
\usepackage{color}
\usepackage[export]{adjustbox} 
\usepackage{floatrow}
\usepackage{float}
\usepackage[font=small,labelfont=bf,justification=Justified]{caption}
\usepackage{subcaption}
\usepackage{makecell}
\usepackage[titles]{tocloft}
\usepackage{titlesec}

\usepackage{verbatim}
\newfloatcommand{capbtabbox}{table}[][\FBwidth]
\usepackage{tabularx}
\usepackage{units}
\usepackage[T1]{fontenc}
\usepackage[normalem]{ulem} 
\usepackage{slashed}
\usepackage{bbold}
\hypersetup{
    colorlinks,
    linkcolor={red!60!black},
    citecolor={blue!50!black},
    urlcolor={blue!80!black}
}
\allowdisplaybreaks

\advance\cftsecnumwidth 0.5em\relax
\advance\cftsubsecindent 0.5em\relax
\advance\cftsubsecnumwidth 0.5em\relax

\newcommand{\rmi}[1]{{\mbox{\scriptsize #1}}}
\newcommand{\rmii}[1]{{\mbox{\tiny\rm{#1}}}}

\newcommand{\MPl}{M_\mathrm{Pl}}
\newcommand{\UBL}{U(1)_\rmii{B-L}}
\newcommand{\mZPrime}{m_\rmii{Z'}}
\newcommand{\gbl}{g_\rmii{B-L}}

\newcommand{\MZ}{m_\rmii{Z}}
\newcommand{\MW}{m_\rmii{W}}
\newcommand{\MT}{m_t}
\newcommand{\Mh}{m_h}

\newcommand{\lamh}{\lambda_h}
\newcommand{\lamphi}{\lambda_\varphi}
\newcommand{\lamp}{\lambda_p}
\newcommand{\BminL}{\mathrm{B-L}}
\newcommand{\csb}{\chi\mathrm{SB}}

\newcommand{\Tc}{T_c}
\newcommand{\Ti}{T_i}
\newcommand{\Tn}{T_n}
\newcommand{\Tp}{T_p}
\newcommand{\Troll}{T_\mathrm{roll}}

\newcommand{\TRH}{T_\rmi{rh}}
\newcommand{\Tqcd}{T_\rmii{QCD}}
\newcommand{\vhqcd}{v_{h,\rmii{QCD}}}
\newcommand{\mQCD}{m_{\varphi,\rmii{QCD}}}

\newcommand{\aRH}{a_\rmi{rh}}
\newcommand{\astar}{a_\star}

\newcommand{\MeV}{\,\mathrm{MeV}}
\newcommand{\GeV}{\,\mathrm{GeV}}

\newcommand{\diff}{{\rm d}}

\newcommand{\Vcw}{V_\rmi{cw}}
\newcommand{\VT}{V_\rmi{T}}
\newcommand{\Veff}{V_\rmi{eff}}
\newcommand{\VDaisy}{V_\rmi{daisy}}
\newcommand{\Vtree}{V_\rmi{tree}}

\titleformat{\section}
  {\centering\normalfont\fontsize{10}{15}\bfseries}{\thesection}{1em}{}
\titleformat{\subsection}
  {\centering\normalfont\fontsize{10}{15}\bfseries}{\thesubsection}{1em}{}

\begin{document}


\title{QCD-sourced tachyonic phase transition in a supercooled Universe}

\author{Daniel Schmitt\,\orcidlink{0000-0003-3369-2253}}
\email{dschmitt@itp.uni-frankfurt.de} 
\affiliation{Institute for Theoretical Physics, Goethe University, 60438 Frankfurt am Main, Germany}

\author{Laura Sagunski\,\orcidlink{0000-0002-3506-3306}}
\email{sagunski@itp.uni-frankfurt.de} 
\affiliation{Institute for Theoretical Physics, Goethe University, 60438 Frankfurt am Main, Germany}

\begin{abstract}
We propose a novel gravitational wave production mechanism in the context of quasi-conformal Standard Model extensions, which provide a way to dynamically generate the electroweak scale. 
In these models, the cosmic thermal history is modified by a substantial period of thermal inflation, potentially supercooling the Universe below the QCD scale. 
The exit from supercooling is typically realized through a strong, first-order phase transition. 
By employing the classically conformal $\UBL$ model as a representative example, we show that a large parameter space exists where bubble percolation is inefficient. 
In this case, the top quark condensate triggers a tachyonic phase transition driven by classical rolling of the new scalar field towards the true vacuum. 
As the field crosses a region where its effective mass is negative, long-wavelength scalar field fluctuations are exponentially amplified, preheating the supercooled Universe. 
We study the dynamics of this scenario and estimate the peak of the associated gravitational wave signal, which is detectable by future observatories in almost the entire available parameter space. 
\end{abstract}

\maketitle
        

\section{Introduction}
Gravitational waves (GWs) are key to studying the largely unknown dynamics of the early Universe.
Since the Standard Model (SM) of particle physics does not predict a significant abundance of relic GWs, the detection of a stochastic gravitational wave background (SGWB) \cite{Caprini:2018mtu} of cosmological origin would profoundly impact our understanding of the fundamental building blocks of nature. 
As a consequence of the high temperatures during the initial stages of the cosmic expansion, future GW experiments such as the Laser Interferometer Space Antenna (LISA)~\cite{2017arXiv170200786A,LISACosmologyWorkingGroup:2022jok} or the Einstein Telescope (ET)~\cite{Punturo:2010zz} provide a way to test beyond the SM (BSM) theories at energies inaccessible to Earth-based colliders.
To be able to deduce the underlying particle theory from an observation, it is therefore important to identify unique signals that emerge in given models.

In this work, we will be concerned with classically conformal (CC) SM extensions~\cite{Meissner:2006zh,Foot:2007iy,Espinosa:2008kw,Iso:2009ss,Iso:2009nw,Iso:2012jn,Farzinnia:2013pga,Englert:2013gz,Hashimoto:2013hta,Khoze:2014xha,Hur:2011sv,Heikinheimo:2013fta,Holthausen:2013ota,Kubo:2014ova,Ametani:2015jla,Kubo:2015joa,Hatanaka:2016rek,Baratella:2018pxi,deBoer:2024jne}.
The common feature of these models is the absence of mass terms in the tree-level potential, providing a mechanism to dynamically generate electroweak symmetry breaking (EWSB) via a new scalar field charged under an additional gauge group.
CC theories offer interesting phenomenology, as they generally predict a substantial period of supercooling, which opens up the window for intriguing baryogenesis~\cite{Konstandin:2011dr,Konstandin:2011ds,Servant:2014bla,Khoze:2013oga,Croon:2019ugf,Azatov:2021irb,Baldes:2021vyz,Huang:2022vkf,Dasgupta:2022isg,Chun:2023ezg} and dark matter (DM)~\cite{Hambye:2013dgv,Carone:2013wla,Khoze:2013uia,Steele:2013fka,Benic:2014aga,Guo:2015lxa,Oda:2017kwl,Hambye:2018qjv,Azatov:2021ifm,Baldes:2021aph,Kawana:2022fum,Khoze:2022nyt,Frandsen:2022klh,Azatov:2024crd} production mechanisms.
The supercooling phase is typically followed by a strong, first-order phase transition (FOPT), generating relic GWs~\cite{Jinno:2016knw,Kubo:2016kpb,Marzola:2017jzl,Iso:2017uuu,Marzo:2018nov,Aoki:2019mlt,Prokopec:2018tnq,Ellis:2020nnr,Wang:2020jrd,Kierkla:2022odc,Sagunski:2023ynd,Kierkla:2023von} in the reach of future observatories.

Employing the CC $\UBL$ model as an example, we show that a detectable FOPT is in fact realized merely in a small part of the available parameter space. 
In the case of extreme supercooling, the Universe keeps inflating as the temperature of the radiation bath approaches the QCD scale, where chiral symmetry breaking ($\csb$)~\cite{Pisarski:1983ms,Brown:1990ev,Braun:2006jd,Cuteri:2021ikv} is triggered.
This breaks the CC symmetry explicitly~\cite{Witten:1980ez,Iso:2017uuu,vonHarling:2017yew,Bodeker:2021mcj,Arunasalam:2017ajm,Ipek:2018lhm,Croon:2019ugf,Sagunski:2023ynd,Guan:2024ccw,Liu:2024fly}, accelerating the end of supercooling. 
We show that this quickly leads to a regime where bubble percolation becomes inefficient.
Then, the top quark condensate sources a tachyonic instability in the BSM scalar field responsible for EWSB.
This induces an exponential amplification of long-wavelength scalar fluctuations, preheating the supercooled Universe.
We will refer to this scenario, where the symmetry breaking is driven by classical rolling instead of thermal tunneling, as \textit{tachyonic phase transition}.\footnote{Note that this scenario is distinct from a conventional second-order phase transition where the scalar field evolves smoothly with the true minimum, hence never experiences a tachyonic instability.}
The growth of scalar fluctuations in a distinct momentum band generates anisotropies in the stress-energy tensor, and hence, GWs~\cite{Dufaux:2007pt,Dufaux:2008dn,Figueroa:2017vfa,Cook:2011hg,Buchmuller:2013lra,Machado:2018nqk,Machado:2019xuc,Figueroa:2020rrl,Bea:2021zol,Madge:2021abk}. 
Tachyonic preheating~\cite{Felder:2000hj,Felder:2001kt,Garcia-Bellido:2001dqy,Copeland:2002ku,Tranberg:2003gi,Desroche:2005yt,Suyama:2006sr,Garcia-Bellido:2007fiu,Dufaux:2007pt,Dufaux:2008dn,Buchmuller:2012wn,Buchmuller:2013dja,Tranberg:2017lrx,Rubio:2019ypq,He:2020ivk, Karam:2021sno,Tomberg:2021bll,Koivunen:2022mem,Dux:2022kuk,Shakya:2023zvs,Brummer:2024ejc} is typically studied in the context of cosmic inflation, where the associated SGWB peaks at frequencies too large to be detectable.
In our scenario, on the other hand, the instability is triggered by QCD, and the peak frequency of the resulting GW spectrum is determined by the CC scale, which we find can be as low as $\mathcal{O}(10^5)\GeV$.
Therefore, we find a large parameter space observable by future experiments.

In sec.~\ref{sec:supercooled_universe}, we introduce the dynamics occurring in CC models.
In the subsequent sec.~\ref{sec:tachyonic_instability}, we identify the parameter space where the tachyonic instability realizes the exit from supercooling, before studying the production of scalar fluctuations analytically and numerically in sec.~\ref{sec:scalar_field_production}.
Afterwards, we describe the reheating of the supercooled thermal bath in sec.~\ref{sec:reheating}. 
Finally, in sec.~\ref{sec:GWs}, we estimate the peak of the associated GW spectrum.

\section{Supercooled Universe} \label{sec:supercooled_universe}
Let us first outline the impact of quasi-conformal dynamics on the thermal history of the Universe.
As a benchmark, we consider the CC $\UBL$ model \cite{Iso:2009nw,Iso:2009ss} where the global $\BminL$ (baryon $-$ lepton number) symmetry of the SM is promoted to a gauge symmetry.
Besides a $Z'$ gauge boson, the model contains an additional, complex scalar field $\Phi = (\varphi + iG)/\sqrt{2}$ with $\BminL$ charge $+2$. Moreover, three right-handed sterile neutrinos are required for anomaly cancellation. These have, however, negligible impact on the dynamics we study and are therefore neglected in the remainder of this work.

The defining feature of the model is scale invariance at tree level, hence the scalar potential reads
\begin{equation} \label{eq:Vtree}
    \Vtree = \lamh H^4 + \lamphi \Phi^4 - \lamp H^2 \Phi^2\, ,
\end{equation}
where $H$ denotes the SM Higgs doublet.
The Higgs and $\BminL$ scalar self-couplings are $\lamh$ and $\lamphi$, respectively, while $\lamp$ is the portal coupling between the electroweak and $\BminL$ sectors. 
Due to the absence of dimensionful terms in the tree-level potential, EWSB is established via the portal term once the $\UBL$ symmetry is spontaneously broken by radiative corrections \cite{Coleman:1973jx}. The vacuum expectation value (VEV) of $\varphi$ can be expressed as
\begin{equation} \label{eq:VEV_definition}
    v_\varphi = \frac{\mZPrime}{2 \gbl} \, ,
\end{equation}
where $\gbl$ is the $\BminL$ gauge coupling and $\mZPrime$ is the $Z'$ boson mass after symmetry breaking.
We note that experimental $Z'$ searches \cite{ATLAS:2017fih,Escudero:2018fwn} dictate $v_\varphi \gg v_h = 246\GeV$. 
Then, the first step of the symmetry breaking pattern is well approximated by merely considering the $\BminL$ direction of the effective potential \cite{Prokopec:2018tnq}. 
In the second step, the EW vacuum is generated by demanding 
\begin{equation} \label{eq:fix_model_params}
    \mu_{h,\rmii{SM}}^2 \simeq \frac{\lamp}{2} v_\varphi^2 \, , \quad \mathrm{and} \quad m_h = 125.10\GeV \,,
\end{equation} 
where $m_h$ is the physical Higgs mass.
This renders $\mZPrime$ and $\gbl$ the only free parameters\footnote{In fact, the model contains a third free parameter, $\Tilde{g}$, which parametrizes the kinetic mixing between the $U(1)_\rmii{Y}$ and $\UBL$ gauge bosons. Its impact on our mechanism is small, hence we set it to $\Tilde{g} = 0$ at the electroweak scale; see appendix~\ref{app:input}.}\footnote{Note that for $\gbl(\mu = m_t) \gtrsim 0.35$, the $\UBL$ gauge coupling runs into a Landau pole~\cite{Khoze:2014xha} below the Planck scale. The parameter space relevant for our mechanism is well below this bound.} in the model (see appendix~\ref{app:input} for more details on the input parameters). 

The scale of EWSB in the early Universe is therefore crucially dependent on the CC dynamics,~i.e., the temperature when $\varphi$ acquires its VEV. 
At high temperatures, thermal corrections restore the $\UBL$ symmetry.
While the Universe cools, the true minimum forms, separated from the origin by a thermal barrier. 
Intriguingly, this barrier remains down to $T \to 0$ as a consequence of classical scale invariance. 
Therefore, tunneling can be significantly delayed, and CC models typically feature strongly supercooled, first-order phase transitions. 
Such transitions are characterized by several temperature scales:
\begin{itemize}
    \item Critical temperature $\Tc$: The VEV becomes degenerate with the minimum at the origin, i.e., tunneling to the true vacuum becomes energetically favorable.
    \item Onset of thermal inflation $\Ti$: If $\varphi$ remains trapped in the false vacuum for a sufficient amount of time, the false vacuum energy starts to dominate the energy budget and the Universe enters a stage of thermal inflation~\cite{Lyth:1995ka,Barreiro:1996dx}.\footnote{Note that this inflationary stage is unrelated to the conjectured initial period of cosmic inflation.}
    \item Percolation temperature $\Tp$: The formation and expansion of bubbles becomes efficient, such that the Universe is converted to the true vacuum state while GW production takes place. 
    \item Reheating temperature $\TRH$: The false vacuum energy is injected back into the thermal bath, which reheats to a temperature $\TRH$. If reheating is fast, we have $\TRH \approx T_i$. Subsequently, the Universe follows the standard evolution.
\end{itemize}

Previous studies \cite{Iso:2017uuu,Marzo:2018nov,Ellis:2020nnr} have shown that for small gauge couplings $\gbl \leq \mathcal{O}(0.1)$, the percolation temperature $\Tp \leq \Tqcd = \mathcal{O}(100\MeV)$. Then, the cosmic QCD transition with six massless flavors occurs first \cite{Pisarski:1983ms,Braun:2006jd,Brown:1990ev,Cuteri:2021ikv} at $\Tqcd \simeq 85\MeV$, breaking the chiral symmetry of QCD via quark condensation, $\expval{q\Bar{q}} \neq 0$. 
At the same time, the finite expectation value of the quark condensate induces a QCD-scale VEV \cite{vonHarling:2017yew,Iso:2017uuu} for the SM Higgs through its Yukawa coupling to the top quark,
\begin{equation}
    \vhqcd = \biggl[-\frac{y_t}{\sqrt{2}\lamh} \langle t\bar{t}\rangle \biggr]^\frac{1}{3}
    \, .
\end{equation}
A precise value of $\vhqcd$ can only be inferred from lattice studies as chiral symmetry breaking ($\csb$) is governed by strongly-coupled dynamics. In addition, $y_t$ runs non-perturbatively large at the QCD scale.  
Following previous studies, we therefore set 
\begin{equation}
    \vhqcd = 100 \MeV\,.
\end{equation}
The QCD-scale Higgs VEV induces a negative mass term in the tree-level potential \eqref{eq:Vtree},
\begin{equation} \label{eq:QCD_induced_mass}
    \Delta \mQCD^2 = -\frac{\lamp}{2} \vhqcd^2  \, ,
\end{equation}
counteracting the thermal barrier. As a consequence, the $\BminL$ PT is either directly sourced or rapidly accelerated such that $\Tp \leq \Tqcd$ \cite{Iso:2017uuu,Marzo:2018nov,Ellis:2020nnr,Sagunski:2023ynd}.

In the following, we demonstrate that the QCD-induced exit from supercooling in CC models is not necessarily realized by a FOPT. Due to the breaking of the CC symmetry by quark condensation, a temperature $\Troll$ emerges where the mass term \eqref{eq:QCD_induced_mass} equals the thermal corrections, $|\Delta \mQCD^2| \sim \gbl^2 \Troll^2$, such that the thermal barrier vanishes. This feature gives rise to a parameter region where the QCD-sourced acceleration of the $\BminL$ PT becomes more and more rapid, such that bubble percolation becomes increasingly inefficient~(see below). Then, $\Tp$ approaches $\Troll$, until these two temperature scales become indistinguishable. In the absence of bubble dynamics, the scalar field $\varphi$ becomes free to roll down the effective potential at $\Troll$, traversing a region in the effective potential where its effective mass is negative, $m_\varphi^2 < 0$. Then, the field experiences a tachyonic instability, leading to an explosive production of long-wavelength scalar fluctuations~\cite{Greene:1997ge}, which preheat the supercooled Universe. In other words, the top quark condensate triggers a tachyonic $\BminL$ phase transition.

\section{QCD-induced tachyonic instability} \label{sec:tachyonic_instability}

{\bf Effective potential.}
Let us now introduce the thermal effective potential $\Veff(\varphi,T)$. The tree-level potential \eqref{eq:Vtree} in the $\varphi$-direction reads
\begin{equation} \label{eq:VtreeQCD}
    \Vtree(\varphi) = \frac{\lamphi}{4} \varphi^4 - \frac{\lamp}{4} \vhqcd^2 \varphi^2 \, ,
\end{equation}
where the QCD term emerges after $\csb$ at $T \simeq 85\MeV$.
Radiative (vacuum) corrections at one-loop order are incorporated as 
\begin{equation} \label{eq:Vcw}
    \Vcw(\varphi,T) = \sum_i \frac{n_i}{64 \pi^2}  M_i(\varphi)^4  \left[\log\left(\frac{M_i(\varphi)^2}{\mu(\varphi,T)^2}\right) - c_i \right]\, ,
\end{equation}
where $n_i$ are the degrees of freedom of a species $i$, $M_i(\phi)$ are the field-dependent masses, and $c_i = \frac{5}{6}\,\left(\frac{3}{2}\right)$ for vector bosons (scalars and fermions). The renormalization group (RG) scale is denoted by $\mu$.

Interactions with the thermal medium generate temperature-dependent corrections to the effective potential of the form
\begin{align} \label{eq:VT}
    \VT(\varphi,T) = \frac{T^4}{2\pi^2}  \sum_i n_i J_\rmii{B,F}\left(\frac{M_i(\varphi)^2}{T^2}\right) \, ,
\end{align}
where
\begin{equation}
    J_\rmii{B,F} (x) = \int_0^\infty {\rm d}y y^2 \ln\left[1 \mp \exp\left(-\sqrt{x+y^2}\right)\right] 
\end{equation}
are the bosonic/fermionic thermal integrals. Thermal resummation is accounted for by implementing the Daisy term~\cite{PhysRevD.45.2933,Arnold:1992rz},
\begin{equation}
    \VDaisy(\varphi,T) = -\frac{T}{12\pi} \sum_i n_i \Big[(M_i(\varphi)^2 + \Pi_i(T))^\frac{3}{2} - M_i(\varphi)^\frac{3}{2}\Big] \, ,
\end{equation}
where $\Pi_i(T)$ denotes the thermal mass of a species $i$, and the sum runs over scalar and longitudinal bosonic degrees of freedom. 
The field-dependent and thermal masses can be found in,~e.g., refs.~\cite{Marzo:2018nov,Ellis:2020nnr}.
Here, we only consider corrections from the $Z'$ gauge boson. 
Since $\lamphi \sim \gbl^4$ (cf. appendix~\ref{app:input}), scalar contributions to $\Veff$ scale as $\sim \gbl^8$, hence, are subleading and will be neglected.
The full effective potential then reads
\begin{equation}
\begin{split}
    \Veff(\varphi,T) = \Vtree&(\varphi) + \Vcw(\varphi,T) \\
    &+ \VT(\varphi,T) + \VDaisy(\varphi,T) \, .
\end{split}
\end{equation}

Note that although a state-of-the-art, dimensionally reduced effective field theory~\cite{Kajantie:1995dw} framework beyond leading order has been established for strongly supercooled phase transitions~\cite{Kierkla:2023von}, we restrict our computation to one-loop order. Since the effective potential suffers from the uncertainty of the strongly coupled dynamics encoded in $\vhqcd$, high-precision thermal resummation plays a subdominant role. 

The RG scale $\mu$ is often chosen as the largest mass scale of the theory, which in our case is $\mZPrime(\varphi)$. 
This, however, becomes ambiguous in theories with a temperature-dependent scale hierarchy. 
If the involved mass scales are small compared to the temperature, the high-temperature approximation holds. 
Then, logarithms involving field-dependent masses in \eqref{eq:Vcw} and \eqref{eq:VT} cancel out~\cite{Gould:2021oba}, leaving a logarithmic dependence on the ratio of $T/\mu$. From this, it becomes clear that the natural choice is a temperature-dependent RG scale $\mu(T)$.\footnote{Typically, RG scales in the range $\frac{\pi}{2} T \leq \mu \leq 4 \pi e^{-\gamma} T$ are employed~\cite{Croon:2020cgk,Lewicki:2024xan}.}
In models that feature strong supercooling, such a treatment only holds in the regime where the field-dependent masses are small. For large field values, e.g., around the true minimum, $\mZPrime(\varphi)/T$ becomes large and the high-$T$ approximation breaks down. Therefore, we follow the approach of ref.~\cite{Kierkla:2023von} and impose\footnote{Note that for energy scales below the QCD scale, $\mu \leq \Lambda_\mathrm{QCD} \simeq 0.1\GeV$, the strong gauge coupling as well as the top quark Yukawa become non-perturbatively large. Hence, we freeze their running below $\Lambda_\mathrm{QCD}$.}
\begin{equation}
    \mu(\varphi,T) = \max \left\{\mZPrime(\varphi), \pi T\right\} \, .
\end{equation}
Then, all couplings that enter the effective potential are evaluated at $\mu(\varphi,T)$. The input parameters are fixed such that the SM vacuum is successfully generated at the electroweak scale $\mu = \MZ$, see app. \ref{app:input}. 
In all figures, we display the $\BminL$ gauge coupling $\gbl$ evaluated at its input scale $\mu = \mZPrime$.\\

{\bf Inefficiency of bubble percolation.}
With the effective potential, we are now able to study the parameter space where bubble nucleation and percolation is (in)efficient. 
We start from a high temperature where the $\UBL$ symmetry is restored and $\varphi$ is trapped in the false vacuum. 
Then, at $\Ti < \Tc$, the Universe enters a period of thermal inflation~\cite{Lyth:1995ka,Barreiro:1996dx} when the false vacuum energy becomes comparable to the energy density in the thermal bath,
\begin{equation} \label{eq:onset_supercooling}
    \rho_r = \frac{\pi^2}{30} g_{\star,\epsilon} T_i^4 = \Delta \Veff(\varphi,T_i) \, ,
\end{equation}
where $\Delta \Veff(\varphi,T)$ is the potential energy difference between the false and true vacuum, and $g_{\star,\epsilon} = 110.75$ denote the effective energetic degrees of freedom of the extended SM. 

The Hubble parameter reads
\begin{equation} \label{eq:Hubble}
    H = \left(\frac{\rho_r(T) + \Delta \Veff}{3 \MPl^2}\right)^\frac{1}{2} \approx \left(\frac{1}{128 \pi^2}\right)^\frac{1}{2} \frac{\mZPrime^2}{\MPl}\ \, ,
\end{equation}
where in the last approximation we have taken the zero-temperature limit of the effective potential.
Hence, $H$ becomes approximately constant for sufficiently small temperatures.
To evaluate the efficiency of bubble nucleation, we compute the decay rate of the false vacuum~\cite{Linde:1981zj}
\begin{equation} \label{eq:tunneling_rate}
    \Gamma(T) \approx T^4 \left(\frac{\mathcal{S}_3}{2\pi}\right)^\frac{3}{2} \exp\left(-\mathcal{S}_3\right) \, .
\end{equation}
Here, $\mathcal{S}_3 = S_3/T$, with the bounce action given by
\begin{equation}
    S_3 \left[\varphi\right] = 4\pi \int \diff r r^2 \left[\frac{1}{2} \left(\frac{\diff\varphi}{\diff r} \right)^2 + \Veff(\varphi,T)\right] \, .
\end{equation}
This expression is evaluated for the bounce solution $\varphi = \varphi_b$, which is obtained by solving the $O(3)$ symmetric equation of motion in Euclidean time,\footnote{To solve the bounce equation, we use a modified version of {\tt CosmoTransitions}~\cite{Wainwright:2011kj}.}
\begin{equation}
    \frac{\diff^2\varphi}{\diff r^2} + \frac{2}{r} \frac{\diff\varphi}{\diff r} = \frac{\diff \Veff (\varphi,T)}{\diff \varphi} \, .
\end{equation}
The nucleation temperature $\Tn$, where bubble nucleation becomes efficient, is defined by~\cite{Linde:1980tt}
\begin{equation}
    \int_{\Tn}^{\Tc} \frac{\diff T}{T} \frac{\Gamma(T)}{H(T)^4} = 1\,,
\end{equation}
corresponding to one nucleated bubble per Hubble volume.
In strongly supercooled phase transitions, however, $\Tn$ is not an appropriate measure for the completion of the PT~\cite{Ellis:2018mja}. Instead, the temperature of bubble percolation $\Tp$ is computed by considering the probability of a point to remain in the false vacuum, $P=\exp(-I(T))$, where~\cite{Guth:1981uk}
\begin{equation}
\label{eq:I(T)}
    I(T) = \frac{4\pi}{3} \int_T^{\Tc}
      \frac{{\diff} T'}{T'^4} \frac{\Gamma(T')}{H(T')}
      \biggl(\int_T^{T'}\!{\diff}\Tilde{T} \frac{v_w}{H(\Tilde{T})}\biggr)^3
      \; .
\end{equation}
The condition for successful percolation reads $I(\Tp) = 0.34$. Note that we choose $v_w = 1$ for the bubble wall velocity that enters eq.~\eqref{eq:I(T)}. This is justified for strongly supercooled phase transitions.

\begin{figure}
    \centering
    \includegraphics[width=\linewidth]{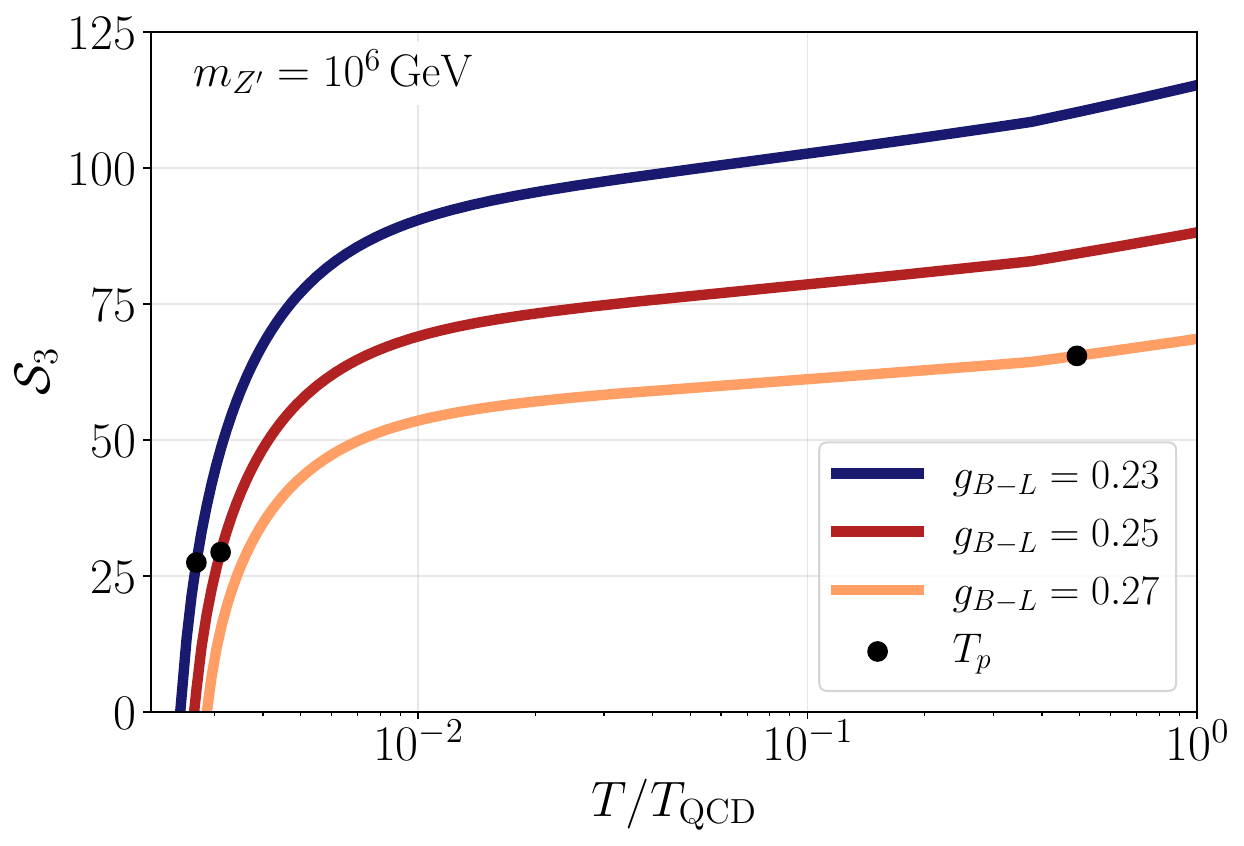}
    \caption{Evolution of the bounce action $\mathcal{S}_3 = S_3/T$ below the QCD scale for $\mZPrime = 10^6\GeV$ and $\gbl \in \{0.23,0.25,0.27\}$. The significant decrease of $\mathcal{S}_3$ is induced by the QCD contribution in $\Veff(\varphi,T)$, which causes the barrier to vanish as $\mathcal{S}_3$ crosses zero at the temperature $\Troll$. The black dots indicate the percolation temperature $\Tp$, which quickly approaches $\Troll$ as the gauge coupling decreases.}
    \label{fig:S3T}
\end{figure}

Fig.~\ref{fig:S3T} shows the evolution of the bounce action $\mathcal{S}_3$ below the QCD temperature for $\mZPrime = 10^6\GeV$ and three different values of the gauge coupling, $\gbl \in \{0.23,0.25,0.27\}$. 
We observe that with decreasing gauge coupling, $\mathcal{S}_3$ initially takes larger values, indicating a stronger suppression of the tunneling rate \eqref{eq:tunneling_rate}. As the temperature decreases, the QCD-induced mass term starts to become comparable to the thermal corrections that generate the potential barrier. 
While the barrier shrinks, the transition is accelerated, i.e., the bounce action decreases. 
The zero-crossing of $\mathcal{S}_3$ marks the temperature $\Troll$ where the barrier vanishes. 

\begin{figure}
    \centering
    \includegraphics[width=\linewidth]{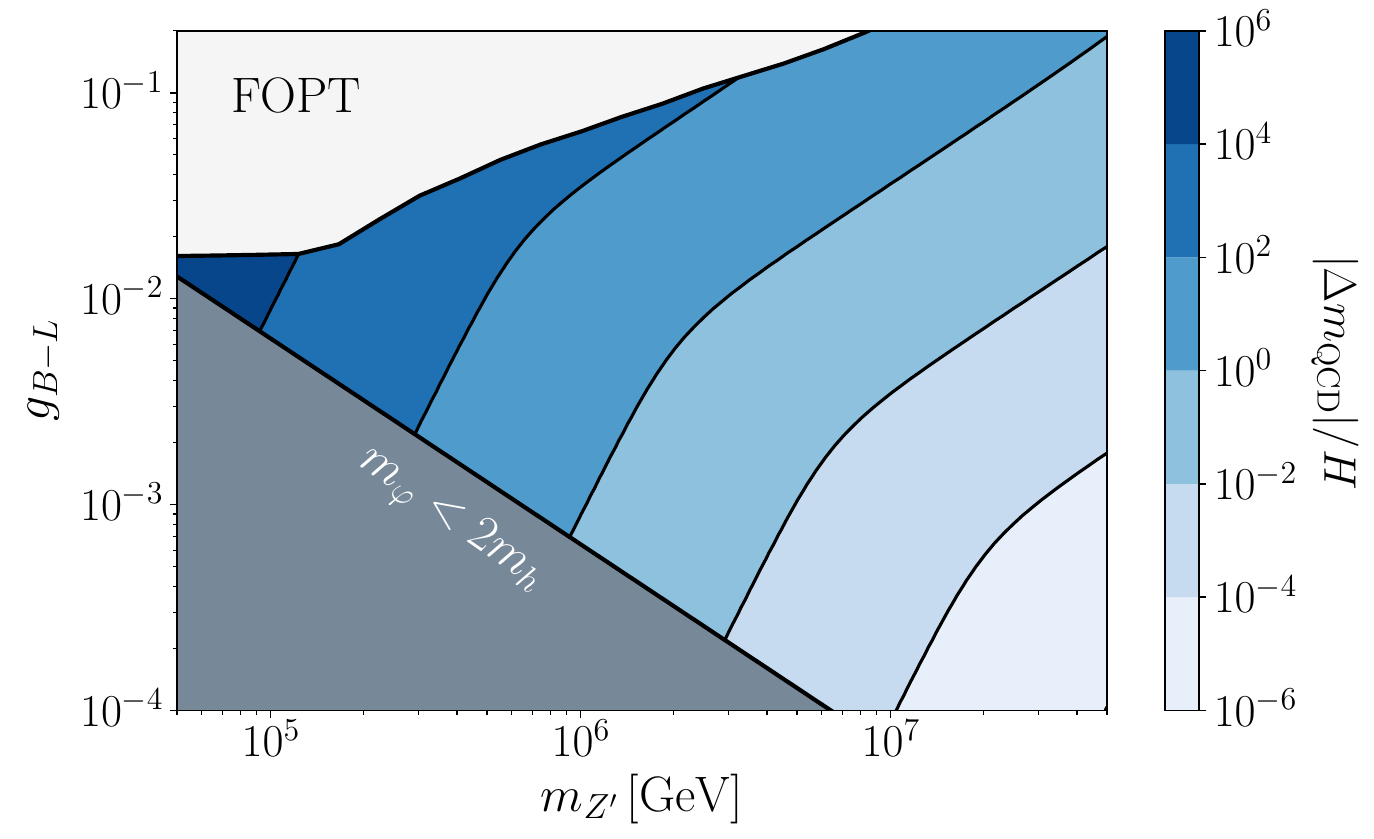}
    \caption{The QCD-induced mass contribution, i.e., the maximal growth rate of scalar fluctuations around the origin of the effective potential, normalized to the Hubble parameter, in the $\mZPrime 
    -\gbl$ plane. The white area indicates where the exit from supercooling is realized by a FOPT (cf.~sec.~\ref{sec:tachyonic_instability}), while the gray region does not allow for successfully reheating the SM (cf.~sec.~\ref{sec:reheating}).}
    \label{fig:mqcd_over_H}
\end{figure}

The black dots in fig.~\ref{fig:S3T} indicate the percolation temperature $\Tp$. 
For smaller values of $\gbl$, $\Tp$ quickly approaches $\Troll$. 
This is due to the stronger initial suppression of $\Gamma(T)$, which requires smaller values of $\mathcal{S}_3$ for successful percolation.
Therefore, we expect that there exists a parameter space where thermal tunneling is not efficient enough to convert the entire Universe to the true vacuum state before the zero-crossing of $\mathcal{S}_3$. 
If we cannot, for given values of ($\gbl,\,\mZPrime$) find a temperature where $I(\Tp) = 0.34$ before the barrier vanishes at $\Troll$, we conclude that the transition becomes second-order. 
As bubble formation is strongly suppressed, the scalar field then remains homogeneous until the barrier disappears at $\Troll$, and subsequently rolls down to the $\UBL$ breaking minimum. Hence, the result is a second-order, tachyonic phase transition, triggered by the change of the mass parameter $m_\varphi^2 = \partial^2\Veff/\partial\varphi^2$ from positive to negative. In fig.~\ref{fig:mqcd_over_H}, we indicate by the white region the parameter space where thermal tunneling drives the transition, while the blue shaded region features a tachyonic transition.

The number of e-folds from the onset of thermal inflation until the barrier vanishes at $\Troll$ is then given by
\begin{equation}
    N = \log\left(\frac{T_i}{\Troll}\right) \,.
\end{equation}
In the parameter space we consider, we find a maximum of $N\sim 25$ e-folds.
In general, a larger value of $\mZPrime$ comes with a more extended supercooling period, as the onset of thermal inflation is shifted to a larger temperature (cf.~eq.~\eqref{eq:onset_supercooling}). 
Also note that $\Troll < T_\rmii{BBN} \sim \mathcal{O}(\mathrm{MeV})$ does not lead to any inconsistencies, as long as the reheating temperature $\TRH$ after thermal inflation is sufficiently large.
\\

\begin{figure}
    \centering
    \includegraphics[width=\linewidth]{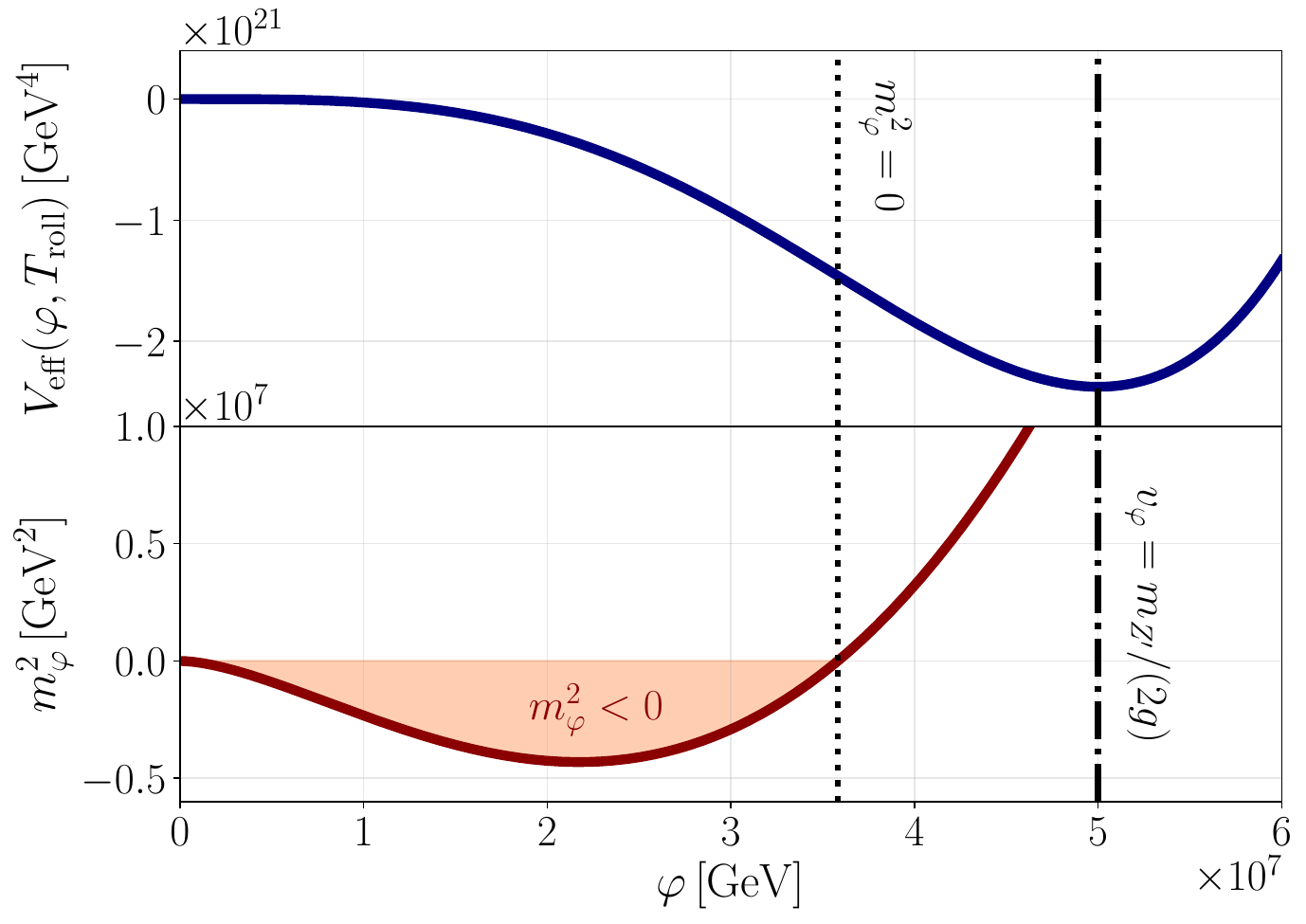}
    \caption{Upper panel: Effective potential at the temperature $\Troll$, where the QCD contribution cancels the thermal barrier, for $\mZPrime = 10^6\GeV$, $\gbl = 10^{-2}$. Lower panel: Associated effective mass squared $\partial^2 \Veff/\partial \varphi^2 = m_\varphi^2$ of the scalar field. Once $\varphi$ starts to roll from small field values, it crosses a region with negative $m_\varphi^2$, corresponding to a tachyonic instability.}
    \label{fig:potential_mass}
\end{figure}

{\bf Tachyonic instability.}
Let us now outline the dynamics of the tachyonic phase transition, which is similar to that in models of tachyonic preheating~\cite{Felder:2000hj,Felder:2001kt,Garcia-Bellido:2001dqy,Copeland:2002ku,Tranberg:2003gi,Garcia-Bellido:2007fiu,Dufaux:2008dn,Tranberg:2017lrx,Rubio:2019ypq,Karam:2021sno,Koivunen:2022mem,Dux:2022kuk,Shakya:2023zvs}.
In these models, the inflaton efficiently preheats the post-inflationary Universe while crossing a tachyonic region of its potential.
Such models are, however, inaccessible to current and future experiments as a consequence of the large energy scale during cosmic inflation.
This is the crucial difference to our work, where the instability is dynamically triggered below the QCD scale, and thus is, excitingly, accessible to experiments.

The equation of motion for the scalar field $\varphi$ in an expanding Universe reads
\begin{equation} \label{eq:scalar_eom}
    \Ddot{\varphi} + \frac{1}{a^2} \nabla^2 \varphi + 3 H \Dot{\varphi} + \frac{\partial \Veff(\varphi,T)}{\partial \varphi} = 0 \, ,\\
\end{equation}
where $a$ denotes the scale factor, $H$ is the Hubble parameter given by eq.~\eqref{eq:Hubble}, and dots indicate derivatives with respect to cosmic time $t$.
As bubble dynamics is absent, the field remains largely homogeneous down to $\Troll$. We can decompose the field into a homogeneous background field $\varphi(t)$ and small fluctuations $\delta \varphi(\boldsymbol{x},t)$,
\begin{equation} \label{eq:field_decomposition}
    \varphi(\boldsymbol{x},t) = \varphi(t) + \delta \varphi(\boldsymbol{x},t) \, .
\end{equation}
In Fourier space, the fluctuations are expressed as~\cite{Dolgov:1989us,Traschen:1990sw,Kofman:1997yn}
\begin{equation} \label{eq:fluctuation_decomposition}
    \delta \varphi(\mathbf{x},t) = \int \frac{{\rm d}^3k}{(2\pi)^{3}} a_k u_{k}(t) \exp(ikt) + h.c.\, ,
\end{equation}
where $a_k^\dagger$, $a_k$ are creation and annihilation operators with
\begin{equation}
    [a_k,a_{k'}^\dagger] = (2\pi)^3 \delta^{(3)}(k-k') \, .
\end{equation}
The mode functions $u_k(t)$ capture the time evolution of a mode carrying momentum $k$.
Inserting eq.~\eqref{eq:field_decomposition} into \eqref{eq:scalar_eom} and keeping only terms up to linear order in $\delta \varphi \ll \varphi$, we obtain equations of motion for the background field and the fluctuations,
\begin{align} \label{eq:eom_decomposed}
    \Ddot{\varphi} + 3 H \Dot{\varphi} + \frac{\partial \Veff(\varphi,T)}{\partial \varphi} &= 0 \, ,\\
    \Ddot{u}_{k} + 3 H \Dot{u}_{k} + \left( \frac{k^2}{a^2} + \frac{\partial^2 \Veff(\varphi,T)}{\partial \varphi^2}  \right)  u_{k} &= 0 \, , \label{eq:eom_fluctuations}
\end{align}
where we have employed eq.~\eqref{eq:fluctuation_decomposition} to express the equation of motion in terms of the mode functions $u_k$.
Hence, a mode with momentum $k$ is described by a harmonic oscillator, damped by Hubble friction, with a time-dependent frequency
\begin{equation} \label{eq:omega_osc}
    \omega^2 = \frac{k^2}{a^2} + \frac{\partial^2 \Veff(\varphi,T)}{\partial \varphi^2} \, .
\end{equation}
For small field values, the scalar mass takes negative values, $\partial^2 \Veff/\partial \varphi^2 = m_\varphi^2 < 0$. 
Then, long-wavelength modes in the range 
\begin{equation} \label{eq:tachyonic_band}
    0 \leq \frac{k}{a} \leq m_\varphi \, ,
\end{equation}
experience a negative effective frequency, $\omega^2 < 0$.
This corresponds to a tachyonic instability, where the solution of eq.~\eqref{eq:eom_fluctuations} changes from an oscillating behavior to exponential growth, $u_k \propto \exp(|\omega| t)$.
Hence, scalar fluctuations are enhanced while $\varphi(t)$ rolls through the concave part of the effective potential.
This is exemplified in fig.~\ref{fig:potential_mass}, where we display $\Veff(\varphi,\Troll)$ for $\mZPrime = 10^6\GeV$, $\gbl = 10^{-2}$, together with the effective mass $m_\varphi^2$ of the scalar field.
The unstable region is indicated by the orange shaded region.

The initial tachyonic band is determined by the effective mass parameter close to the origin,
\begin{equation} \label{eq:massOrigin}
    m_{\varphi,0}^2 = \frac{\partial^2\Veff(\varphi,\Tqcd)}{\partial \varphi^2}\Big|_{\varphi \to 0} \simeq \Delta \mQCD^2 + \gbl^2 T^2 \, ,
\end{equation}
where we have only kept leading-order terms. As the temperature decreases below $\Troll$, thermal effects quickly become negligible, such that the effective mass is dictated by the negative, QCD-induced contribution~\eqref{eq:QCD_induced_mass}. 

We display the maximal initial growth rate $\omega \simeq |\Delta \mQCD|$, divided by the Hubble parameter \eqref{eq:Hubble}, in fig.~\ref{fig:mqcd_over_H}. 
We observe that for large VEVs, i.e., large $\mZPrime$ and small $\gbl$, the effective potential flattens, hence the effective mass and thus the growth rate decreases. 
Vice versa, the effective mass term increases for smaller VEVs.
Then, $|\mQCD| > H$, and the tachyonic instability becomes effective right at $\Troll$.

\section{Scalar field amplification} \label{sec:scalar_field_production}
{\bf Initial conditions.}
To study the exponential growth of scalar fluctuations, we first need to specify the initial conditions.
Prior to $\Troll$, the homogeneous component $\varphi(t)$ is trapped in the false vacuum around the origin, while the Universe is exponentially expanding.
At the end of thermal inflation, the expectation value of the scalar field is dictated by quantum fluctuations that evolve during the inflationary period.
These may be split into a thermal $\langle \varphi^2 \rangle_\rmii{T} \approx T^2/12$ and a vacuum $\langle \varphi^2 \rangle_\rmii{v} \approx H^2 N/(4\pi^2)$ part~\cite{VILENKIN1983527,Dimopoulos:2019wew,Lewicki:2021xku}, such that $\langle \varphi^2 \rangle = \langle \varphi^2 \rangle_\rmii{T} + \langle \varphi^2 \rangle_\rmii{v}$.
Classical rolling eventually takes over when~\cite{Hook:2014uia}
\begin{equation}
    \Veff'(\varphi_\mathrm{cl}) = -\frac{3H^3}{2\pi} \, .
\end{equation}
Therefore we need to ensure the classical limit is justified.
First, this includes points where $\varphi_\mathrm{cl} \leq \sqrt{\langle \varphi^2 \rangle}$ is fulfilled at $\Troll$.
In the regime where $\sqrt{\langle \varphi^2 \rangle} < \varphi_\mathrm{cl}$, but $|\mQCD| \gg H$, the growth of quantum fluctuations typically drives the field towards the classical regime rather quickly~\cite{Lewicki:2021xku}.
For the parameter space where $|\mQCD| \ll H$, however, thermal inflation is expected to continue for a sizable number of e-folds after $\Troll$.
This may lead to contradictions with CMB observations~\cite{Planck:2018jri}, or even to eternal inflation~\cite{Dimopoulos:2019wew}.
Since an analysis of the quantum fluctuations during thermal inflation is beyond the scope of this work, we focus on the parameter space where $|\mQCD| > H$, and choose
\begin{equation}
    \varphi_i = \max\Big\{\varphi_\mathrm{cl}, \sqrt{\langle \varphi^2 \rangle}\Big\}
\end{equation}
as initial condition for the zero-mode.\footnote{Note that tachyonic growth is however not very sensitive on the initial condition, as the field spends most time in the regime where the effective mass parameter $\Delta m_\varphi^2 \approx \Delta \mQCD^2 = \mathrm{const.}$, i.e., the growth rate is largely independent of the zero-mode.}

While the homogeneous mode is trapped in the false vacuum, the scalar field is effectively massless before QCD confinement, and the subhorizon scalar modes are interpreted as relativistic particles which carry momentum $k$. In cosmic preheating scenarios, the initial condition for the mode functions is then given by the Bunch-Davies vacuum~\cite{Birrell:1982ix}
\begin{equation} \label{eq:bunch_davies}
    u_{k,\rmii{BD}}(\eta) = \frac{1}{\sqrt{2k}} \exp(i k \eta) \, ,
\end{equation}
where $\eta$ denotes conformal time. 
In our case, we have to take into account the presence of the thermal bath. 
Assuming the scalar fluctuations to be in equilibrium with the SM,\footnote{Since the scalar fluctuations are amplified by many orders of magnitude, this assumption has no significant impact on the dynamics. We have explicitly checked this by comparing our results to the ones obtained with Bunch-Davies initial conditions.} their associated energy density reads
\begin{align} \label{eq:spectral_energy_dens_maxwell_boltzmann}
    \rho_{\delta \varphi} (T) &= \int {\rm d}k \, \frac{k^3}{2 \pi^2}  \left[{\exp \left(\frac{k}{T}\right) - 1} \right]^{-1}\\
    &= \frac{\pi^2}{30} T^4\, ,
\end{align}
where we integrate over physical momenta.
On the other hand, we may express the scalar fluctuation energy density as
\begin{align} \label{eq:energy_density_fluctuations}
    \rho_{\delta\varphi} = \frac{1}{2} \delta \dot{\varphi}^2 + \frac{1}{2a^2}(\nabla \delta \varphi)^2 \, .
\end{align}
From this, we find appropriate initial conditions for the mode functions, 
\begin{equation}
    u_k(\eta) = \sqrt{2} \left( \exp \left(\displaystyle \frac{k}{\Troll}\right) - 1\right)^{-\frac{1}{2}} u_{k,\rmii{BD}}(\eta)\, .
\end{equation}

To conclude, let us comment on the evolution of the SM Higgs field, which initially sits at $\vhqcd = \mathcal{O}(\Tqcd) \sim 100\,\MeV$.
As $\varphi$ rolls towards its true minimum, the curvature in the Higgs direction becomes negative at some point, generating the electroweak vacuum.
However, we have checked that this occurs at times well beyond the initial phase of tachyonic amplification.
Therefore, we can safely neglect the Higgs dynamics in our study.\\

\begin{figure}
    \centering
    \includegraphics[width=\linewidth]{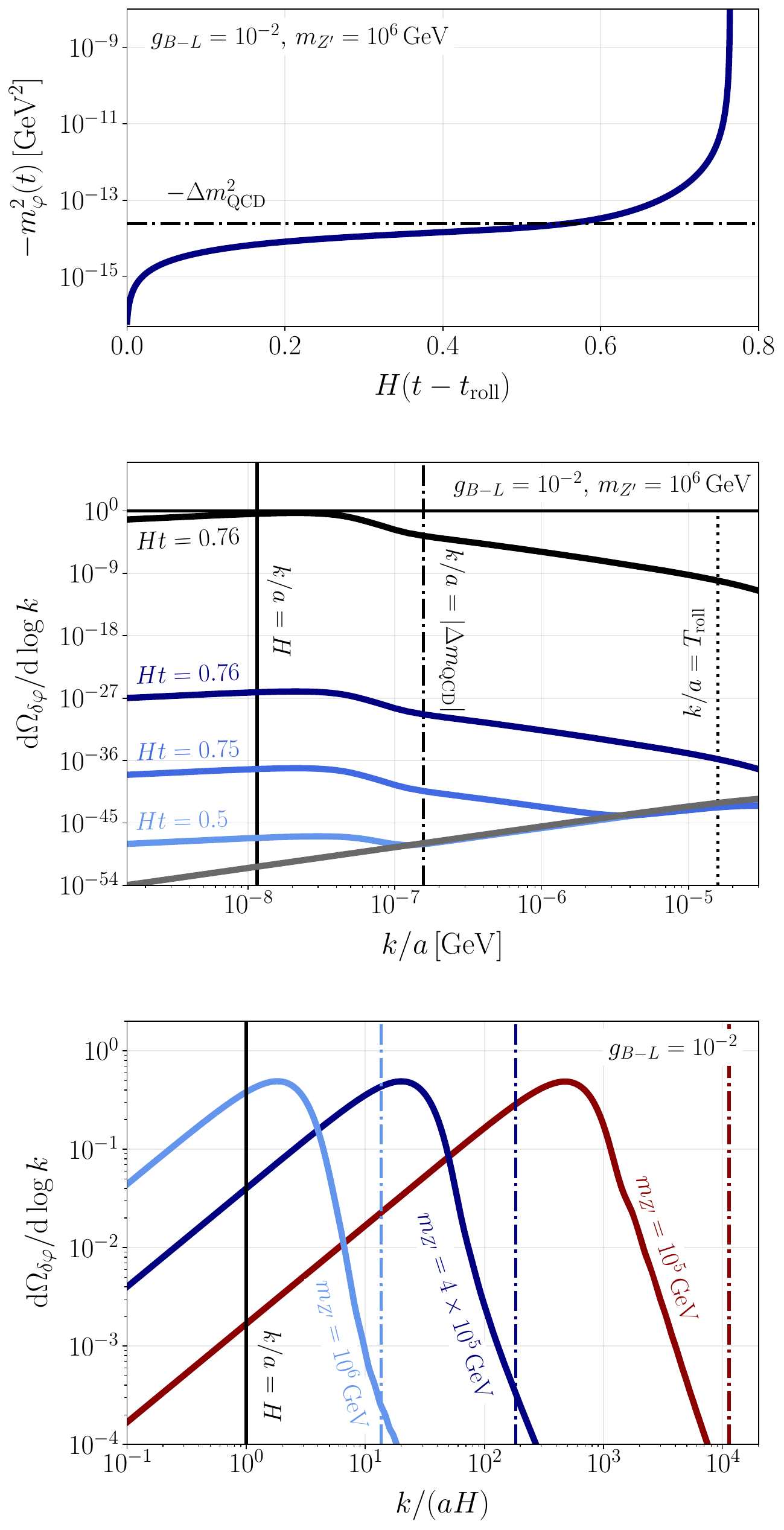}
    \caption{Results from our numerical simulation of the tachyonic instability. Upper panel: negative effective scalar mass parameter as a function of time. Center panel: energy spectrum of the amplified fluctuations at different times, normalized to the total energy density. Lower panel: final scalar fluctuation energy spectrum for different benchmark parameters. The dash-dotted lines indicate the respective QCD-induced mass terms, controlling the width of the instability band.}
    \label{fig:scalar_evolution}
\end{figure}

{\bf Numerical study.}
With the above initial conditions, we are now set to study the exponential production of scalar fluctuations. 
Note, however, that the linearized equations of motion \eqref{eq:eom_decomposed} do not capture the backreaction from the the excited modes onto the zero-mode. 
As the energy density of the fluctuations becomes comparable to the one of the zero-mode, fragmentation sets in. 
Then the field configuration becomes fully inhomogeneous, and the scale of inhomogeneity is given by the peak momentum of the amplified modes.
Such non-linear dynamics can only be studied using dedicated lattice simulations \cite{Figueroa:2020rrl}.
Nevertheless, the linearized analysis provides an intuition about the efficiency of tachyonic growth. 
In addition, we obtain a reliable prediction of the peak momentum, which we will employ in sec.~\ref{sec:GWs} to estimate the peak of the associated GW spectrum.

In our linearized analysis, we solve 
eqs.~\eqref{eq:eom_decomposed} and \eqref{eq:eom_fluctuations} numerically for $N_k = 1000$ momentum modes.
In the absence of non-linear effects, we then stop the simulation as the energy density of the fluctuations, $\rho_{\delta \varphi}$, approaches the total initial energy density of the scalar, $\rho_\mathrm{tot} = \Delta V$.
To compute the energy density of the scalar excitations, we first note that the occupation number of a harmonic oscillator with frequency $\omega_k$ reads
\begin{equation}\label{eq:particle_number}
    n_k = \frac{\omega_k}{2} \left(\frac{|\dot{u}_k|^2}{\omega_k^2} + |u_k|^2 \right) - \frac{1}{2}\, .
\end{equation}
The associated energy density is then
\begin{equation}
    \rho_{\delta\varphi} = \int \frac{{\rm d}^3 k}{(2\pi)^3} \omega_k \left(n_k + \frac{1}{2}\right) \, .
\end{equation}
Note that eq.~\eqref{eq:particle_number} formally only holds in the adiabatic limit $\dot{\omega}_k/\omega_k^2 \ll 1$.
This condition is violated once $\varphi$ reaches larger field values, where the second derivative of $\Veff$, i.e., the growth rate $\omega_k^2$, rapidly increases.
Also, the concept of a particle number density is ill-defined while the mass is tachyonic, as $\omega_k^2 < 0$ for the unstable modes.
However, the main goal of our numerical study is to verify that the peak momentum scale is indeed dictated by the QCD-induced mass.
We will see that this peak forms during the early stage of the evolution where the adiabaticity condition holds. 
A more careful treatment of the particle number will therefore not alter the location of the peak, but merely the time when tachyonic growth ends.
Once the scalar field rolls down to its minimum, $m_\varphi^2$ becomes positive and the particle interpretation becomes valid.
In the tachyonic regime, the energy density is then typically computed~\cite{Felder:2001kt} by defining either
\begin{equation} \label{eq:omega_k_choices}
    \omega_k = \biggr|\frac{k}{a}\biggr|\,, \quad \mathrm{or} \quad \omega_k = \sqrt{\left(\frac{k}{a}\right)^2 + |m_\varphi^2|} \, ,
\end{equation}
The impact of the choice of $\omega_k$ on our final results proves to be small, hence we employ $\omega_k = |k/a|$ in the remainder of this work.

Fig.~\ref{fig:scalar_evolution} shows the result of our numerical analysis.
In the upper panel, we display the negative effective mass parameter as a function of time for the benchmark parameters $\mZPrime = 10^6\GeV$ and $\gbl = 10^{-2}$.
Initially, at $\Troll$, the QCD-induced mass is balanced by thermal corrections, hence $m_\varphi^2 = 0$ around the origin.
As the temperature decreases and the field starts to roll close to the origin, its effective mass approaches $-\Delta \mQCD^2$, indicated by the dash-dotted line. 
Subsequently, $\varphi$ quickly rolls towards larger field values, where the potential becomes dominated by the radiative corrections. 
Then, the effective mass rapidly becomes more negative, i.e., the growth rate increases by several orders of magnitude.

In the center panel, we plot the corresponding 
fluctuation energy spectrum. 
Different colors denote different times, where the gray curve indicates the thermal spectrum at the onset of rolling, $t_\mathrm{roll} = 0$, which peaks at $k/a \sim \Troll$. 
Initially, the growth rate is dictated by the QCD-induced mass term marked by the dash-dotted line. 
In this regime, only modes with $k/a < |\Delta \mQCD|$ grow efficiently.
As the effective mass becomes more negative at $Ht \approx 0.6$ (cf. upper panel), modes with larger momentum enter the instability band.
Since then the growth rate $\omega \sim |m_\varphi| \gg H$, energy transfer becomes increasingly efficient.
Hence, the total initial energy density is transferred into fluctuations within less than one Hubble time, before oscillations around the true minimum start. 
Let us stress that due to the quasi-conformal nature of the effective potential, the zero mode spends a significant time close to the origin.
Therefore, the cutoff and peak of the final spectrum at the end of the simulation, i.e., where $\rho_{\delta \varphi} = \rho_\varphi$, are still dictated by $\Delta \mQCD^2$, as those low-momentum modes are amplified during the entire evolution.

In the lower panel, we show the final spectra for different benchmark parameters, with $\mZPrime \in \{10^5, 4 \times 10^5 ,10^6\}\GeV$ and $\gbl = 10^{-2}$.
Here, we normalize the momenta to the Hubble parameter. 
The dash-dotted lines again indicate the QCD-induced mass, which determines the characteristic momentum scale for all benchmarks.
Note, however, that the peak of the spectrum is generally located below $|\Delta \mQCD|$, as modes with momenta close to $k/a \lesssim |\Delta \mQCD|$ experience a smaller growth rate (cf.~eq.~\eqref{eq:eom_fluctuations}).
For smaller $\mZPrime$, this effect becomes more prominent, leading to an $\mathcal{O}(1-10)$ overestimation of the peak scale.
For our purposes, however, $\Delta \mQCD$ will be a sufficiently accurate estimate of the peak momentum.

\section{Reheating} \label{sec:reheating}
To ensure a consistent cosmological evolution, the energy of the excited scalar modes has to be efficiently transferred back to the SM sector.
After the tachyonic transition, the scalar field is driven to the global minimum of the effective potential, corresponding to the $\BminL$ and electroweak vacuum, $(\langle\varphi\rangle,\langle h\rangle) = (v_\varphi, 246\GeV)$.
Then all SM fields are rendered massive.
The scalar fluctuations which dominate the energy density of the Universe may now be interpreted as particles with mass
\begin{equation}\label{eq:scalar_mass}
    m_\varphi = \sqrt{\frac{\partial^2 \Veff(\varphi,T)}{\partial \varphi^2}}\Biggr|_{\varphi=v_\varphi} \approx \sqrt{\frac{6}{\pi^2}} \gbl^2 v_\varphi  \, .
\end{equation}
Here, we have taken the zero-temperature limit which is justified as $v_\varphi \gg \Troll$. 
In addition, we have neglected scalar mixing because of the small mixing angle $\lamp \ll 1$.
For simplicity, we focus on reheating via the Higgs portal term in eq.~\eqref{eq:Vtree}, hence we require $m_\varphi > 2 m_h$.\footnote{Note that scalar mixing may provide a successful reheating mechanism \cite{Kawai:2023dac} if $m_\varphi < 2\Mh$.}
The decay rate of $\varphi \to hh$ reads~\cite{Ellis:2020nnr,Kawai:2023dac}
\begin{equation}\label{eq:decay_rate}
    \Gamma_\varphi \approx \frac{\lamp^2 v_\varphi^2}{32 \pi m_\varphi} \, ,
\end{equation}
where we evaluate the model parameters at the EW scale, $\mu = \MZ$.
In most of the parameter space, this expression is smaller than the expansion rate of the Universe, $\Gamma < H$, prohibiting the scalar decay initially.
Since the scalar modes carry physical momenta $k \ll m_\varphi$, hence are non-relativistic, the Universe undergoes a matter-domination period until
\begin{equation}\label{eq:H_rh}
    H_\mathrm{rh} = H_\star \left(\frac{a_\star}{a_\mathrm{rh}}\right)^\frac{3}{2} = \Gamma_\varphi \, ,
\end{equation}
where $H_\star$ and $a_\star$ denote the Hubble parameter and scale factor at the end of tachyonic growth. 
Assuming a quick thermalization of the SM bath, the reheating temperature is approximated by
\begin{equation}
    \frac{\pi^2}{30} g_{\star,\epsilon} T_\mathrm{rh}^4 = 3 \MPl^2 \mathrm{min}\left(H_\star^2, \Gamma_\varphi^2\right) \, .
\end{equation}
To provide some intuition, we can for now neglect the running of the parameters. Combining eqs.~\eqref{eq:Hubble}, \eqref{eq:scalar_mass}, and \eqref{eq:decay_rate}, we then obtain
\begin{equation}
    \frac{\Gamma_\varphi}{H_\star} \approx 2\times 10^{-3} \gbl \left(\frac{10^6\GeV}{\mZPrime}\right)^5 \, ,
\end{equation}
in agreement with ref.~\cite{Ellis:2020nnr}.
Hence, small gauge couplings and large $Z'$ masses lead to a more extended period of matter domination.
This yields
\begin{equation}
    \TRH \approx 1.4 \times 10^4  g_{\star,\epsilon}^{-\frac{1}{4}} \gbl^\frac{1}{2}  \left(\frac{10^6\GeV}{\mZPrime}\right)^\frac{3}{2} \GeV \, ,
\end{equation}
where $g_{\star,\epsilon}$ denotes the energetic degrees of freedom in the thermal bath after reheating.

From this, we can derive an approximate bound such that the reheating temperature is larger than the temperature of big bang nucleosynthesis, $T_\rmii{BBN} \sim \mathcal{O}(\mathrm{MeV})$,
\begin{equation}
    \mZPrime < 5.8\times 10^{10} g_{\star,\epsilon}^{-\frac{1}{6}} \gbl^\frac{1}{3} \GeV\,.
\end{equation}
This is well outside the parameter space we consider.

\section{Gravitational waves} \label{sec:GWs}
\begin{figure*}
    \centering
    \includegraphics[width=\textwidth]{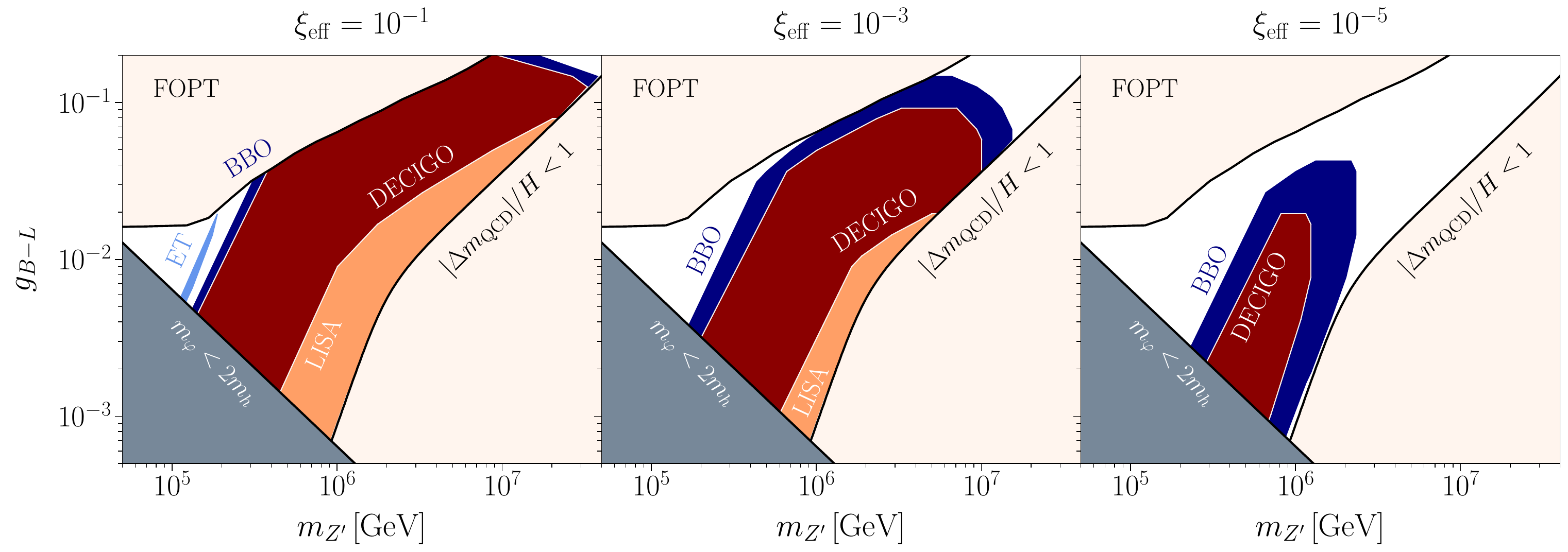}
    \caption{Projected sensitivities of future GW experiments based on our estimate of the gravitational wave peak~\eqref{eq:OmegaGW_peak}, for three different values of the efficiency factor $\xi_\mathrm{eff}$. For large $\xi_\mathrm{eff}$, BBO and DECIGO will be able to probe almost the entire parameter space. Particularly strong signals are in the reach of LISA, while ET only covers a small part. In the upper left exclusion band the exit from supercooling is triggered by a FOPT instead of a tachyonic transition. The gray shaded region indicates the regime where $\varphi$ cannot reheat the Universe by decays to the Higgs. In the right exclusion region, the scalar mass around the origin is smaller than the Hubble parameter, i.e., the assumption of classical rolling does not hold.}
    \label{fig:gw_prospects}
\end{figure*}

The spectral energy density of a stochastic GW background $d\rho_\rmii{GW}(f)/d\log f$, normalized to the total energy density $\rho_\rmi{tot}$, reads
\begin{equation}
    \Omega_\rmii{GW} (f) \equiv \frac{1}{\rho_\rmi{tot}} \frac{d\rho_\rmii{GW}(f)}{d\log f} \, , \quad \rho_\rmii{GW} = \frac{\MPl^2}{4}\langle{\dot{h}_{ij} \dot{h}^{ij}}\rangle \; .
\end{equation}
The evolution of the metric perturbations $h_{ij}$ associated with the growth of scalar field fluctuations is obtained by solving the linearized Einstein equations
\begin{equation} \label{eq:einstein_eqs}
    h''_{ij} + 2\mathcal{H} h'_{ij} - \nabla^2 h_{ij} = \frac{2}{\MPl^2} \Pi_{ij}^\rmii{TT} \, .
\end{equation}
Here, primes denote derivatives with respect to conformal time $\eta$, $\mathcal{H} = a'/a$ is the comoving Hubble parameter, and $\Pi_{ij}^\rmii{TT}$ is the transverse-traceless part of the stress-energy tensor of the system. In our model, GW production occurs during several stages of the evolution. 

The first source is given by the exponential amplification of scalar fluctuations in a distinct momentum band during the linear stage of tachyonic growth.
To this end, one solves eq.~\eqref{eq:einstein_eqs} by plugging in the scalar energy-momentum tensor to obtain an expression that depends directly on the gradient energy of the scalar field fluctuations; see ref.~\cite{Figueroa:2017vfa} for the full computation.
Note that, however, although the initial total energy is efficiently injected into scalar fluctuations, we find negligible GW emission from the linear stage, as the potential energy is predominantly converted into kinetic fluctuations.
This is in agreement with previous studies of tachyonic preheating in hybrid inflation models: 
in refs.~\cite{Dufaux:2007pt,Dufaux:2008dn}, it was shown using lattice simulations that the significant part of GW production takes place when the system becomes non-linear.
In this phase, bubble-like structures form, expand, and eventually collide, which induces large field gradients that in turn generate GWs. 
Since a lattice study is beyond the scope of this work, we restrict ourselves to order-of-magnitude estimates and relegate a more precise computation to the future.

According to~\cite{Dufaux:2007pt,Dufaux:2008dn}, the GW peak amplitude in tachyonic preheating models is well approximated by
\begin{equation}\label{eq:OmegaGW_peak}
    \Omega_\rmii{GW} \approx \xi_\mathrm{eff}(R_\star H_\star)^2 \, .
\end{equation}
Here $R_\star \sim a/k_\star$ is the size of the induced inhomogeneities which sets the peak frequency. 
The efficiency of GW production is determined by the model-dependent pre-factor $\xi_\mathrm{eff}$.
In the case of chaotic inflation~\cite{Dufaux:2007pt}, e.g., $\xi_\mathrm{eff} \sim 0.15$.
In our case, we will treat $\xi_\mathrm{eff}$ as a free parameter.

Eq.~\eqref{eq:OmegaGW_peak} is familiar from FOPTs, where the peak amplitude $\Omega_\rmii{GW} \propto (H_\star/\beta)^2$ is suppressed by the inverse timescale of the transition $\beta$.
For strongly supercooled transitions, $\xi_\mathrm{eff} \sim \mathcal{O}(10^{-2})$~\cite{Lewicki:2022pdb}.
Since the spectral shape is however expected to differ, our scenario would be distinguishable from a FOPT in the case of a detection.

In sec.~\ref{sec:scalar_field_production}, we verified that the QCD-induced mass term indeed sets the cutoff scale of the scalar spectral energy density, however, generally overestimates the peak momentum. 
Nevertheless, we will employ $|\Delta \mQCD|$ as characteristic scale, as we are merely interested in an order-of-magnitude estimate of the GW peak.
Also, an overestimation of $k_\star$ corresponds to an underestimation of the peak amplitude, i.e., our following results are to be understood as conservative estimates.
Regarding the peak momentum at the time of production, we obtain
\begin{equation}
    \begin{split}
        \frac{k_\star^2}{a^2} &\approx \frac{\lamp}{2} \vhqcd^2 \\
        &\approx 4 \mu_{h,\rmii{SM}}^2 \vhqcd^2 \frac{\gbl^2}{\mZPrime^2} 
    \; ,
    \end{split}
\end{equation}
where we have neglected the running of $\lamp$ in the second step. 
In addition, we have not incorporated the effect from the expansion of space from the onset of rolling to the time of GW production.
This is justified since tachyonic growth completes within much less than a Hubble time in most of the parameter space.
The Hubble parameter during thermal inflation is given by eq.~\eqref{eq:Hubble}.
Today's amplitude and frequency are then obtained by redshifting.
Taking into account the matter domination period, we have~\cite{Kamionkowski:1993fg}
\begin{align}
    &f_0 = 1.65 \times 10^{-7} \mathrm{Hz} \frac{k_\star}{a H_\mathrm{rh}} \frac{\TRH}{\mathrm{GeV}} \left(\frac{g_{\star,\mathrm{rh}}}{100}\right)^\frac{1}{6} \frac{\astar}{\aRH}\,, \\
    &h^2 \Omega_{0,\rmii{GW}} = 1.67\times 10^{-5} \left(\frac{100}{g_{\star,\mathrm{rh}}}\right)^\frac{1}{3} \frac{\astar}{\aRH} \Omega_{\star,\rmii{GW}}\,,
\end{align}
where we have set $g_{\star,s} = g_{\star,\epsilon}$ at the time of reheating.
The ratio of scale factors $a_\star/a_\mathrm{rh}$ is read off from eq.~\eqref{eq:H_rh}.
This can now be expressed in terms of our model parameters.
We find, for $|\Delta \mQCD| > H$,
\begin{widetext}
\begin{align}
    f_0 &\approx 
    \begin{cases}
         \displaystyle 0.57 \, g_{\star,\mathrm{rh}}^{-\frac{1}{12}} \left(\frac{\gbl}{10^{-2}}\right)^\frac{7}{6}  \left(\frac{\vhqcd}{\mathrm{GeV}} \right) \left(\frac{10^6\,\mathrm{GeV}}{\mZPrime}\right)^{\frac{17}{6}} \,\mathrm{Hz} \, , & \text{if $\Gamma_\varphi < H_\star$}\, ,\\
        \displaystyle 23.2 \,g_{\star,\mathrm{rh}}^{-\frac{1}{12}} \left(\frac{\gbl}{10^{-2}}\right)^\frac{1}{2} \left(\frac{\vhqcd}{\mathrm{GeV}} \right)\left(\frac{\mZPrime}{10^3\,\mathrm{GeV}}\right)^\frac{1}{2} \, \mathrm{Hz}\, , & \text{if $\Gamma_\varphi \geq H_\star$}\, ,
    \end{cases} \\
    h^2 \Omega_{0,\rmii{GW}} &\approx 
    \begin{cases}
        \displaystyle  5.5\times 10^{-13} \,\xi_\mathrm{eff} \left(\frac{100}{g_{\star,\mathrm{rh}}}\right)^\frac{1}{3} \left(\frac{10^{-2}}{\gbl}\right)^\frac{4}{3}\left(\frac{\mathrm{GeV}}{\vhqcd} \right)^2 \left(\frac{\mZPrime}{10^6\GeV}\right)^\frac{8}{3} \, , & \text{if $\Gamma_\varphi < H_\star$}\, , \\
        \displaystyle 7.0\times 10^{-10}\,\xi_\mathrm{eff} \left(\frac{100}{g_{\star,\mathrm{rh}}}\right)^\frac{1}{3} \left(\frac{10^{-2}}{\gbl}\right)^2 \left(\frac{\mathrm{GeV}}{\vhqcd}\right)^2 \left(\frac{\mZPrime}{10^6\GeV}\right)^6 \, , & \text{if $\Gamma_\varphi \geq H_\star$}\, .
    \end{cases}
\end{align}
\end{widetext}
Therefore, a larger QCD-induced VEV $\vhqcd$ of the Higgs field shifts the GW signal to a higher frequency and smaller amplitude.
The same applies to the gauge coupling $\gbl$.
This can be directly seen from eq.~\eqref{eq:QCD_induced_mass}, as $(k_\star/a)^2 \approx \Delta \mQCD^2 \propto (\gbl\vhqcd)^2$.
The impact of the gauge boson mass is a bit more subtle. 
If reheating is fast, larger $\mZPrime$ moves the peak towards larger frequencies. 
This is expected as $\mZPrime$ sets the Hubble parameter during thermal inflation.
For slow decays of the scalar field and hence long reheating periods, this behavior changes and large $\mZPrime$ decrease the peak frequency. 
This is due to the strong suppression of the scalar decay rate, which leads to an extended redshift $a_\star/a_\mathrm{rh} \propto \mZPrime^{-10/3}$.
Regarding the peak amplitude, larger gauge boson masses generally lead to stronger signals, albeit the scaling of $h^2 \Omega_\rmii{GW}$ changes depending on the duration of the reheating process.

Let us stress again that for the above estimates, we neglected the running of the model parameters. 
For our final results, we evaluate the decay rate at the global minimum characterized by $\MZ$, and use the thermal RG scale $\mu \sim \pi \Troll$ to compute the QCD-induced mass relevant for tachyonic growth.
In fig.~\ref{fig:gw_prospects}, we present our final results for $\xi_\mathrm{eff} \in \{10^{-1}, 10^{-3}, 10^{-5}\}$. 
Here, the exclusion areas correspond to the parameter space where either the exit from supercooling is realized by a FOPT, the $\BminL$ scalar is too light to successfully reheat the Universe, or the scalar mass around the origin is smaller than the Hubble parameter, potentially causing eternal inflation.
The colored areas are in the reach of the future observatories ET~\cite{Punturo:2010zz} (light blue), BBO~\cite{Crowder:2005nr} (blue), DECIGO~\cite{Seto:2001qf,Sato:2017dkf,Kawamura:2020pcg} (red), and LISA~\cite{2017arXiv170200786A,LISACosmologyWorkingGroup:2022jok} (orange).
To this end, we compute the power-law integrated sensitivity curves $h^2\Omega_\rmii{PLI}(f)$ for the respective experiments.
We then only include points where our estimate of the peak $h^2 \Omega_{0,\rmii{GW}}(f_0) \geq h^2\Omega_\rmii{PLI}(f_0)$.
While this is a simplified approach, we obtain an informative overview of the expected sensitivities.
In general, we find the strongest GW signals near the excluded region where $|\Delta \mQCD| \leq H$, as the scale of induced fluctuations moves closer to the size of the horizon, $|\Delta \mQCD| \sim H$.
This implies great observational prospects at BBO and DECIGO, which will be able to probe almost the entire parameter space if $\xi_\mathrm{eff}$ is sufficiently large. 
Moving towards smaller $\mZPrime$ and larger $\gbl$, the peak is shifted to larger frequencies.
At the same time, the amplitude is suppressed since $|\Delta \mQCD| \gg H$~(see fig.~\ref{fig:mqcd_over_H}), which is why LISA only covers a smaller part of the overall area.
For the same reason, it is difficult to find GW signals accessible by ET, apart from the tiny blue shaded region in the left panel.

\section{Conclusions}
We have presented a mechanism to end the supercooling period characteristic to classically scale-invariant SM extensions that has not been addressed in the literature so far, leading to a unique GW signal. 
While previous works have mostly analyzed the exit from thermal inflation via a strong first-order phase transition, we identify, employing the $\UBL$ model, a large parameter space where bubble percolation is inefficient.
The breaking of classical scale invariance by the top quark condensate instead induces a cancellation of the thermal barrier, and the $\UBL$ scalar becomes free to roll down its effective potential towards the true minimum.
The field crosses a tachyonic instability, as its effective mass parameter is negative close to the origin.
This leads to a copious production of sub-horizon scalar field fluctuations, which quickly dominate the energy density of the Universe.
These fluctuations emerge as scalar particles after the $\UBL$ symmetry is broken and efficiently reheat the supercooled Universe by decaying to Higgs bosons.
Intriguingly, the exponential amplification of scalar fluctuations induces a sizable SGWB.
We have estimated the peak frequency and amplitude of the GW spectrum based on previous lattice results~\cite{Dufaux:2007pt,Dufaux:2008dn}.
The characteristic frequency scale is given by the interplay between QCD dynamics and mass of the new gauge boson, $\mZPrime \sim \mathcal{O}(10^5-10^7)\GeV$, which dictates the onset of thermal inflation.
Therefore, we find promising observational prospects at the future observatories LISA, DECIGO, ET, and BBO.
This is distinct from previous works where tachyonic preheating follows cosmic inflation.
The high scale of inflation typically shifts the signal to the high-frequency band, which is not accessible in the near future. 

Finally, let us comment on possible next directions. 
First, our analysis may be extended by higher loop orders in the effective potential.
This allows for a more reliable prediction of the parameter space where thermal tunneling is inefficient, i.e., where the tachyonic instability becomes relevant.
To this end, a better understanding of the QCD-induced vacuum expectation value of the Higgs field is required, as well.
This can, however, only be inferred from lattice studies of massless QCD augmented by the Higgs.
Regarding the tachyonic instability, we have restricted ourselves to the linear regime. 
This limits our study of the associated GW generation to order-of-magnitude estimates.
A classical lattice simulation, taking into account all non-linear effects, is therefore needed to obtain a more robust prediction of the GW signal, in particular of the spectral behavior.
Then it would be interesting to study the ability of future experiments to distinguish between the tachyonic phase transition and the standard scenario of a supercooled FOPT.
We will return to these exciting questions in the future.

\begin{acknowledgments}
We acknowledge enlightening discussions with D.~Bödeker, 
T.~Caldas Cifuentes,
M.~Kierkla, 
J.~Kuß,
M.~Lewicki,
N.~Ramberg,
P.~Schicho, 
P.~Schwaller,
B.~\'Swie\.zewska,
and
J.~van de Vis during various stages of this work.
We also acknowledge support by
the Deutsche Forschungsgemeinschaft (DFG, German Research Foundation) through
the CRC-TR 211 `Strong-interaction matter under extreme conditions' --
project no.\ 315477589 -- TRR 211.
\end{acknowledgments}

\appendix
\section{Input parameters}\label{app:input}
In this section, we briefly describe how to determine the input parameters, following ref.~\cite{Kierkla:2022odc}.
The model introduces five unknown quantities: the scalar self-coupling $\lamphi$, the portal coupling $\lamp$, the gauge coupling $\gbl$, the scalar VEV $v_\varphi$, and the kinetic mixing $\Tilde{g}$ between the $\UBL$ and $U(1)_\rmii{Y}$ gauge groups. 
The scalar self- and portal coupling, however, are fixed by the requirement to generate the EW scale. In addition, we express our results in terms of $\mZPrime$ instead of $v_\varphi$.
The procedure then is:
\begin{enumerate}[(i)]
    \item The scalar VEV at $\mu = \mZPrime$ is obtained from the tree-level relation $\mZPrime = 2 \gbl v_\varphi$. We then demand that loop corrections preserve the VEV, 
    \begin{equation}
        \frac{{\rm d} \Vcw}{{\rm d}\varphi}\biggr|_{\varphi = v_\varphi,\mu=\mZPrime} = 0 \, ,
    \end{equation}
    which fixes
    \begin{equation}
        \lamphi(\mu = \mZPrime) = \frac{\gbl^4}{\pi^2} \, .
    \end{equation}
    This is a typical relation in Coleman-Weinberg models.

    \item Using the one-loop $\beta$-functions~\cite{Khoze:2014xha}, we run $\lamphi$ and $\gbl$ down to the EW scale $\mu=\MZ$. From the physical masses $\Mh$, $\MZ$, $\MW$, and $\MT$, we obtain the SM gauge couplings and the top quark Yukawa. Since $v_\varphi \gg v_h$ in all of the parameter space we consider, we neglect mass mixing between the two scalars. Then also $\lamh$ takes its SM value.
    
    \item We minimize the $\BminL$ potential for $\mu = \MZ$ to find $v_\varphi(\mu=\MZ)$. The portal coupling is then determined by demanding that $h$ acquires its SM tree-level VEV $v_{h,\rmii{SM}}\simeq 246\GeV$. This translates to
    \begin{equation}
        \lamp(\mu = \MZ) = 2 \lamh \left(\frac{v_{h,\rmii{SM}}}{v_\varphi}\right)^2 \, .
    \end{equation}

    \item This fixes all input parameters at $\mu = \MZ$. For the computation of $\Veff(\varphi,T)$, we then use the $\beta$-functions to run the parameters to the respective scale
    \begin{equation}
        \mu = \max\{\mZPrime(\varphi),\pi T\} \, ,
    \end{equation}
    as outlined in the main body.
\end{enumerate}
The kinetic mixing parameter $\Tilde{g}$ only enters our analysis via the RG evolution.
In previous studies \cite{Marzo:2018nov,Ellis:2020nnr}, it was shown that choosing $\Tilde{g}(\mu=\mZPrime) = -0.5$ renders the EW vacuum stable up to the Planck scale.
However, we find that this leads to numerical instabilities in the running of $\lamp$ at low energy scales $\mu \sim \Tqcd$, where tachyonic growth takes place.
Therefore, we choose kinetic mixing to vanish at the EW scale, $\Tilde{g}(\mu=\MZ) = 0$, to minimize its impact on our results.

\bibliographystyle{apsrev4-1}
\bibliography{biblio.bib}
\end{document}